# Estimating multidimensional poverty under incomplete data in North Macedonia

Maja Gerovska Mitev[1] and Viktor Stojkoski[2,*]

[1] Ss. Cyril and Methodius University in Skopje, Faculty of Philosophy, Institute of Social Work and Social Policy, Skopje, North Macedonia. ORCID: 0000-0003-3854-5199

[2] Ss. Cyril and Methodius University in Skopje, Faculty of Economics-Skopje, Skopje, North Macedonia. ORCID: 0000-0002-7594-4431

[*] Corresponding author: Viktor Stojkoski, e-mail: vstojkoski@eccf.ukim.edu.mk

## Abstract

Multidimensional Poverty Indices (MPIs) are available for more than 100 countries, but rarely below the national or regional level. Local estimates require knowing how deprivations overlap within households, yet the sources that cover every locality, such as censuses and geospatial data, cannot be linked at the household level. We show how to measure local multidimensional poverty under such incomplete data, using North Macedonia as a case study. Combining settlement-level access to services with municipal census data, we pair an independence-based point estimate with sharp bounds over all joint distributions consistent with the data. We find that, nationally, 5.01 percent of households are multidimensionally poor, with bounds of 2.01 to 17.44 percent. Also, poverty is concentrated in small rural municipalities, which are poorer than nearly all municipalities of the capital city under any dependence structure. Validating our index against official statistics, the Global MPI, and a fieldwork deprivation index indicates that the socio-economic part of the index is consistent with external measures of deprivation, while the geographic part captures a distinct dimension of remoteness. Our approach thus presents a promising way to measure multidimensional poverty at the local level in countries where linked household data are unavailable.

## 1 Introduction

Multidimensional measures have become a standard tool for poverty measurement over the last two decades. Drawing on Sen's capability approach (Sen, 1985), they assess the disadvantages that people experience in several aspects of life. The most widely used is the Multidimensional Poverty Index (MPI), based on the counting approach of Alkire and Foster (2011a). The Global MPI uses this approach to measure acute poverty in health, education, and living standards in more than 100 countries (Alkire and Santos, 2014; Alkire et al., 2024).

For local policy, however, we need estimates for individual municipalities and settlements that household surveys rarely provide. Indeed, national and regional averages can conceal substantial differences between these places, while decisions about schools, health facilities, and basic infrastructure depend on local conditions.

For social policy, a national headcount is of limited use, as it does not show where deprivation is located or whom it affects. Social spending, institutional generosity and poverty outcomes often change in different directions (Sowula et al., 2024), and national averages hide the places where deprivation is concentrated. Hence, a national figure alone cannot support decisions on targeting.

To fill this gap, local poverty measures often draw on censuses. Their detailed geographic coverage has long supported measures of unmet basic needs (Feres and Mancero, 2001). Colombia, for example, calculates a municipal MPI directly from census indicators recorded for the same households (DANE, 2020). Census data can also be supplemented with survey information to model missing indicators, as in Mexico (CONEVAL, 2021). Nevertheless, both approaches still require household-level records. Moreover, the survey-based approach also requires compatible survey data, which may be outdated or unavailable (Kovesdi and Alkire, 2026).

To calculate an MPI from these sources, we need to know how deprivations overlap within households (Alkire et al., 2015). A census may describe housing and education in detail, while service records and geographic data provide additional information on access to facilities. If these sources cannot be linked to the same households, we cannot directly observe how their deprivations overlap (Ferreira and Lugo, 2013). Two municipalities can have the same rates of low education and poor housing but different poverty rates, depending on whether these deprivations affect the same households or different ones. Area-based indices can combine separate local indicators (McLennan et al., 2019), but they do not recover this household overlap.

North Macedonia illustrates the problem well. The most recent survey suitable for multidimensional poverty measurement is the 2018–2019 Multiple Indicator Cluster Survey (MICS), on which the OPHI Global MPI reports a national headcount of only 0.4 percent and estimates for eight statistical regions (OPHI, 2024). Also, the Survey on Income and Living Conditions cannot be linked to census records at household level. No estimates exist for the country's 80 municipalities, which are responsible for the provision of key local services. Yet a wealth of local information is available. The 2021 Census reports detailed education, employment, and housing conditions for every municipality, and the national road network and facility registers allow us to measure access to kindergartens, schools, and health care for every settlement in the country. The challenge is to combine these sources into a multidimensional poverty measure without observing how their deprivations overlap.

In this paper, we address this challenge and measure multidimensional poverty for all 80 municipalities and eight regions of North Macedonia under incomplete data. We construct a thirteen-indicator household MPI from two sources. The first contains six indicators of geographic access to basic services, which we compute jointly for every settlement in the country. The second contains seven census indicators of education, employment, and housing, reported as municipal marginal rates in the tabulations we use. Following Brunckhorst et al. (2026), we compute a point estimate by assuming independence wherever the joint distribution is unobserved. We then use linear programming to find the lowest and highest values of the poverty headcount and the MPI that are consistent with everything we do observe.

We find a national poverty headcount of 5.01 percent under independence (MPI = 0.0219), with bounds of 2.01 to 17.44 percent. The municipal point estimates range from nearly zero in several urban municipalities of the capital to almost two-thirds in some small rural municipalities. We also find that some of these differences hold under every joint distribution consistent with the data. The regional bounds, however, overlap, so we cannot establish a regional poverty ranking without additional assumptions.

To interpret these estimates, we examine which deprivations are most common and how our index relates to external evidence. We find that distance to a kindergarten, low adult educational attainment, and distance to a secondary school are the most common deprivations. In comparisons with official demographic and economic statistics and a fieldwork deprivation index, the socio-economic sub-index shows stronger associations than the geographic sub-index.

## 2 Multidimensional poverty and the limits of local data

Household surveys are the main source for measuring multidimensional poverty, as they record all deprivations for the same households. Their samples, however, are designed to represent countries and large regions, and are usually too small to describe individual municipalities. Censuses, administrative registers, and geographic data cover every locality, but each records only some of the relevant deprivations. In addition, administrative geographic data are often of poor quality (Nájera Catalán, 2018). As a result, local measurement requires combining sources, which raises two questions: how to estimate each indicator locally, and how to recover the way indicators overlap within households.

The first question has a long tradition in poverty mapping. Small-area estimation predicts welfare at a fine geographic scale using characteristics recorded in both a household survey and a census (Elbers et al., 2003; Molina and Rao, 2010). Its key assumption is that the relationships estimated in the survey also hold in the places being predicted. When they do not, the resulting estimates can be biased and their confidence intervals too narrow (Tarozzi and Deaton, 2009). In Uganda, for example, this method produced the first sub-county poverty maps (Uganda Bureau of Statistics and UNICEF, 2019). However, it requires enough variables measured in the same way in both sources, as well as access to census microdata, which is not the case in North Macedonia. More recently, satellite imagery and machine learning have expanded the set of predictors (Jean et al., 2016), although with larger errors in the poorest areas (Corral et al., 2025). These methods extend the geographic coverage of a welfare measure, but leave the second question open.

The second question is the domain of statistical matching, which combines sources that do not jointly observe all variables of interest (D'Orazio et al., 2006; Ridder and Moffitt, 2007). Such sources are typically consistent with many joint distributions, each implying a different poverty estimate. A point estimate selects one of them by assumption, whereas partial identification reports the full range of outcomes consistent with the data (Manski, 2003). Brunckhorst et al. (2026) bring this approach to multidimensional poverty, pairing an estimate based on independence with bounds over all feasible joint distributions. Their framework preserves the relationships that are observed and makes explicit how much the remaining ones matter. Wide bounds are informative in their own right. They show what the available data can settle, and whether some places remain poorer than others across the whole range.

How the missing relationships are filled in matters, because deprivations tend to affect the same people. Rippin (2010) and Decancq (2014) show how this concentration shapes poverty and welfare comparisons. Empirically, deprivation indicators are strongly associated (Alkire and Ballon, 2012), a few joint profiles account for most of the world's multidimensionally poor (Suppa et al., 2022), disadvantages accumulate and persist over time (Suppa, 2025), and the network of co-deprivations is broadly stable across countries (Stojkoski et al., 2025). Independence is thus more likely to spread deprivations across households than to reproduce their true concentration. These patterns, however, are specific to particular indicators and scales, and must be established anew in each setting. This is why Ferreira and Lugo (2013) call for analyzing the joint distribution of deprivations in addition to their marginal rates, and why the debate between Ravallion (2011) and Alkire and Foster (2011b) stresses that an index should be reported with its components.

The choice of dimensions is a further question. The income-poor and the deprivation-poor are different groups (Menshikov et al., 2020). Anand et al. (2021) find that capability, income and subjective-wellbeing poverty identify the same people as poor in no more than half of the cases. Consequently, the choice of dimensions determines who is counted as poor.

A second layer of choices concerns the poverty measure itself. The axiomatic and counting literatures establish the properties of alternative identification and aggregation rules (Tsui, 2002; Atkinson, 2003;

Bourguignon and Chakravarty, 2003; Alkire and Foster, 2011a; Bossert et al., 2013). Applying any of them requires choosing indicators, weights, and cutoffs, and these choices can change both poverty levels and rankings (Decancq and Lugo, 2013; Azpitarte et al., 2020; Carrino et al., 2024). Duclos et al. (2006) show how to test whether poverty comparisons hold across aggregation procedures and poverty lines. With incomplete data, this sensitivity adds to the uncertainty about the missing relationships, since a ranking can be stable across weights and still depend on how deprivations overlap. A credible local measure should therefore examine both.

Geographic data can also add new dimensions to a poverty measure. Maps of travel time to services reveal large differences between places that broad-area averages conceal (Weiss et al., 2018). Road distance captures one aspect of accessibility, alongside transport, travel time, individual circumstances, and service quality (Geurs and van Wee, 2004). Adding access indicators thus broadens what a poverty measure describes, from the conditions inside a household to the service environment around it.

This broader scope shapes how a local measure should be compared with existing indices. Measures of acute poverty, of monetary and non-monetary deprivation, and of child poverty differ in their indicators and populations (Alkire and Santos, 2014; World Bank, 2018; Diaz-Bonilla et al., 2022; de Neubourg et al., 2013), as do the European measures that complement income-based assessments (Nolan and Whelan, 2010; Whelan et al., 2014; Suppa, 2018). Within a consistent specification, geographic and group decompositions allow direct comparisons (Alkire and Seth, 2015; Alkire et al., 2024). Across differently designed indices, agreement is most informative for the dimensions they share. We follow this principle in our validation, comparing each part of our index with the external measures that capture similar deprivations.

We contribute to this literature by showing what incomplete local data can establish about multidimensional poverty. Building on Brunckhorst et al. (2026), we combine municipal census indicators with settlement-level access to services, preserving the observed overlap among geographic deprivations. We then identify the municipal comparisons that hold across all joint distributions consistent with the data, and examine their sensitivity to weights, distance thresholds, and indicators (Duclos et al., 2006; Decancq and Lugo, 2013). Considering both sources of uncertainty separates the conclusions that existing data support from those that require further data or assumptions.

For North Macedonia, we provide, to our knowledge, the first MPI covering all 80 municipalities and combining household living conditions with access to services. It complements existing evidence on child poverty, including in Roma settlements (Srbinoski et al., 2023), on rural poverty (Tuna et al., 2022), and on acute poverty at the national and regional level (OPHI, 2024).

## 3 Data and methods

We measure multidimensional poverty for North Macedonia's 80 municipalities and eight statistical regions by combining two data collection efforts. The first covers geographic access to basic services, which we compute for each of the country's 1,575 settlements using the national road network. The second covers socio-economic living conditions, which we draw from municipal tabulations of the 2021 Population Census. Together, they cover 598,632 households. We use the household as our unit of analysis and weight the aggregates by settlement household counts.

### 3.1 Data

**Geographic access to services.** Our first dataset measures how far households must travel to reach basic public services. We select six destinations: i) kindergarten, ii) primary school, iii) secondary school, iv) primary health-care facility, v) hospital, and vi) public park or green space.

For each settlement, we compute the shortest road-network distance to the nearest facility of each type, using a drivable road-network graph of the country extracted from OpenStreetMap. Locations of schools and hospitals come from the national open data portal, and locations of primary health-care facilities from the Health Insurance Fund's register of primary-care physicians. For primary schools, we also include 173 satellite schools, which provide primary education in smaller settlements, and use the shorter of the routes to a main or a satellite school. Parks and green spaces come from OpenStreetMap's tagged features. We cross-check settlement coordinates and education locations against the 2021 Census, OpenStreetMap, and manual review. The exact locations of six education facilities could not be confirmed and are approximate.

We flag a settlement as deprived when this distance exceeds a threshold (Table 1). We use 2.5 km for primary schools, roughly three to five times the catchment radius set by the Ministry of Education and Science's 2020 norms, and extend the same threshold to kindergartens, for which no national norm exists. We use 5 km for secondary schools, twice the distance at which the Law on Secondary Education entitles students to free transport. We use 10 km for primary health care and 30 km for hospitals, twice the distances that the 2025 Rulebook on Health Care Protection uses to classify a settlement as rural. For green space, where no national norm exists either, we use 10 km.

All households in a settlement are assigned the settlement's deprivation status, including for kindergartens and schools, regardless of whether the household has children. In this sense, the geographic indicators measure the service environment in which a household lives, rather than its immediate demand for a particular service. Because we compute all six distances for the same settlements, the geographic dataset is the one part of our data where we directly observe the joint distribution of deprivations. We know, for instance, whether a settlement is far from both a school and a hospital, and not only how common each condition is on its own.

**Table 1. Geographic access indicators, deprivation thresholds, and weights.**

| Indicator | Deprivation threshold | Indicator weight |
|---|---|---|
| Kindergarten access | Nearest kindergarten more than 2.5 km by road | 1/12 |
| Primary school access | Nearest primary or satellite school more than 2.5 km by road | 1/12 |
| Secondary school access | Nearest secondary school more than 5 km by road | 1/12 |
| Primary health-care access | Nearest primary health-care facility more than 10 km by road | 1/12 |
| Hospital access | Nearest hospital more than 30 km by road | 1/12 |
| Green-space access | Nearest park or green space more than 10 km by road | 1/12 |

Note: Distances are computed for settlements and assigned to all their households. The geographic domain has a total weight of one half.

**Socio-economic living conditions.** Our second dataset measures socio-economic living conditions through seven census indicators covering education and employment, and housing (Table 2).

The three education and employment indicators are household-level rules. A household is deprived in low education if any member aged 25–64 has at most primary education, following the ISCED 2011 definition used by EU-SILC. Early school leaving applies the same rule to members aged 18–24. Since the census records attainment rather than current enrollment, this indicator is a proxy for early school leaving. NEET applies if any member aged 15–29 is not employed. We keep the census label, but note that, unlike the standard definition, it does not require the person to also be outside education or training.

We define the four housing indicators directly from the census categories. A household is deprived in sanitation if it has no toilet or the toilet is outside the dwelling, and in housing if it lives in a building of weak or temporary construction, another inadequate housing type, or is homeless. It is deprived in heating if it reports no heating of any kind, and in water if its source is anything other than piped water from the public network or its own pump.

Five of the indicators come from cross-tabulations produced by the State Statistical Office, and the remaining two, water and heating, from published census tables. Each gives the number of households per municipality that meet the condition, which we divide by the municipal household total. We harmonize municipality names and treat empty source categories as zero during preprocessing. The census tables made available to us give municipal counts for each condition but do not identify which households experience it. Consequently, we observe only the marginal rate of each indicator, never their joint distribution with each other or with the geographic indicators.

**Table 2. Socio-economic (census) indicators, deprivation definitions, and weights.**

| Indicator | Household deprivation condition | Indicator weight |
|---|---|---|
| Low education | At least one member aged 25–64 with at most primary education | 1/14 |
| Early school leaving | At least one member aged 18–24 with at most primary education | 1/14 |
| NEET | At least one member aged 15–29 who is not employed | 1/14 |
| Inadequate sanitation | No toilet or toilet outside the dwelling | 1/14 |
| Dilapidated housing | Weak or temporary construction, other inadequate housing category, or homelessness | 1/14 |
| No heating | No heating of any kind | 1/14 |
| No piped water | Neither public-network nor own-pump piped water in the dwelling | 1/14 |

Note: Rates are computed over all households in each municipality. The socio-economic domain has a total weight of one half. NEET follows the census definition and does not require the person to be outside education or training.

## 3.2 Methods

We measure multidimensional poverty using the adjusted headcount ratio of Alkire and Foster (2011a). Let a household's deprivation pattern be described by a geographic vector g with six binary entries and a socio-economic vector z with seven, where one denotes deprivation. Following the standard nested convention (Alkire and Santos, 2014), each domain receives a weight of one half, divided equally among its indicators. This gives 1/12 to each geographic indicator and 1/14 to each socio-economic indicator. The household's deprivation score is then

$$c(g,z) = \frac{1}{12}\sum_{r=1}^{6} g_r + \frac{1}{14}\sum_{j=1}^{7} z_j \,. \quad (1)$$

A household is multidimensionally poor when its score exceeds the poverty cutoff k = 1/3. Let $b_k(g,z)$ equal one in that case and zero otherwise, and let $\pi_m(g,z)$ be the share of households in municipality m with pattern (g,z). The headcount ratio H and the adjusted headcount ratio $M_0$, which we refer to as the MPI, are

$$H_m(\pi_m;\ k) = \sum_{g,z} \pi_m(g,z) b_k(g,z) \ ;$$

$$M_{0m}(\pi_m;\ k) = \sum_{g,z} \pi_m(g,z) c(g,z) b_k(g,z)\ . \quad (2)$$

H is the share of poor households, while the MPI also accounts for the intensity of their deprivation, A, since $M_0 = H \times A$. To illustrate the cutoff, a household deprived in three geographic and two socio-economic indicators has a score of 0.393 and is poor, as is a household deprived in five socio-economic indicators alone (0.357). By contrast, a household deprived in two geographic and two socio-economic indicators scores 0.310 and is not poor. A household deprived in exactly four geographic indicators scores exactly one third, and since we require the score to exceed the cutoff, it is not poor either. We examine the sensitivity of our results to this choice in Appendix A.

Equation (2) makes the challenge we face explicit. Computing H and the MPI requires the full joint distribution $\pi_m(g,z)$, which has 64 geographic times 128 socio-economic patterns, or 8,192 cells per municipality. Our data provide the geographic part of this distribution and the seven socio-economic margins, but not how they combine. We address this in three steps, following Brunckhorst et al. (2026).

**Step 1: A point estimate under independence.** Let $q_m(g)$ be the share of households in municipality m living in settlements with geographic pattern g, and let $p_{mj}$ be the observed deprivation rate of socio-economic indicator j. The standard way to fill in an unobserved joint distribution is to assume conditional independence given a variable that is observed in both sources. Beyond municipality, no common household-level covariates are available to model the missing relationships. In the absence of such covariates, we assume that the socio-economic indicators are independent of each other and of the geographic pattern within each municipality. The share of households with any complete pattern is then a simple product:

$$\pi_m^I(g,z) = q_m(g) \prod_{j=1}^{7} {p_{mj}}^{z_j} (1 - p_{mj})^{1-z_j}\ . \quad (3)$$

This distribution preserves the observed joint distribution of the geographic indicators and reproduces every census margin. Since municipal rates differ, it does not assume independence across the country as a whole. We evaluate H and the MPI by exact enumeration of all cells in each municipality.

**Step 2: Bounds via linear programming.** The independence assumption selects one of many joint distributions that are consistent with our data. As discussed in Section 2, there is good reason to expect that deprivations cluster rather than occur independently. To show how much our results depend on this assumption, we compute the lowest and highest values of H and the MPI over all distributions consistent with the data. A distribution is feasible if its cells are non-negative, it reproduces the geographic distribution, and it reproduces the census margins:

$$\Pi_m = \{\pi_m \geq 0 : \sum_{z} \pi_m(g,z) = q_m(g), \forall g\ ;$$

$$\sum_{g,z} z_j \pi_m(g,z) = p_{mj}, \forall j\ \}. \quad (4)$$

For each statistic $T \in \{H, M_0\}$, the lower and upper bounds are

$$T_m^L = min_{\pi_m \in \Pi_m} T_m(\pi_m;\ k)\ ;\ \ T_m^U = max_{\pi_m \in \Pi_m} T_m(\pi_m;\ k)\ . \quad (5)$$

Since both objectives are linear in the cell probabilities, we solve four linear programs per municipality, minimizing and maximizing H and the MPI, using the HiGHS solver (Huangfu and Hall, 2018). Together with the cutoff sensitivity analysis, this amounts to 1,280 linear programs, all of which solve to optimality. For a single pair of indicators with deprivation rates p and q, the share of households deprived in both can lie anywhere between max(0, p + q − 1) and min(p, q). These limits are known as the Fréchet–Hoeffding bounds, and they are a standard way of describing the uncertainty that arises when sources are combined without a common link (D'Orazio et al., 2006). Our linear programs apply the same logic to the full thirteen-indicator distribution. The resulting bounds are sharp: each endpoint is attained by a feasible distribution, so no narrower interval can be justified by the data alone. Note that the extreme values of H and the MPI need not be attained by the same distribution, so the bounds should be read separately for each statistic.

**Step 3: Aggregation to regions and the country.** Each municipality's bounds come from a separate linear program that uses only its own data, and no constraint links one municipality to another. Hence, every municipality can attain its own minimum, or maximum, at the same time. This allows us to aggregate by simple household-weighted averages, where $\omega_m$ is municipality m's share of households in the region or country:

$$T^I = \sum_m \omega_m T_m^I \; ; \quad T^L = \sum_m \omega_m T_m^L \; ; \quad T^U = \sum_m \omega_m T_m^U \, . \qquad (6)$$

Altogether, this gives point estimates and sharp bounds for 80 municipalities, eight regions, and the country as a whole.

## 4 Results

### 4.1 National and regional poverty

At the national level, we obtain a point estimate of H = 5.01 percent (MPI = 0.0219), and poor households are on average deprived in 43.7 percent of the weighted indicators. The bounds span 2.01 to 17.44 percent for H and 0.0089 to 0.0696 for the MPI. Hence, the observed data alone are consistent with a headcount that is less than half or more than three times as large as the point estimate. The point estimate also sits much closer to the lower bound than to the upper one.

We present the regional results in Figure 1A–B. Southeast has the highest regional MPI (0.0370, H = 8.61 percent), more than four times that of the capital region, Skopje, which has the lowest (0.0082, H = 1.91 percent). Since the capital region is home to 32.2 percent of the country's households, its low value pulls down the national average considerably. Conversely, the three most deprived regions are also the three smallest by population.

The regional bounds, however, overlap. Even the lower MPI bound of the most deprived region (0.0116) is below the upper bound of the capital region (0.0395), and the bounds of every other pair of regions overlap as well. Hence, the regional ranking holds under independence, but the data alone cannot rule out a dependence structure that would reverse it.

### 4.2 Municipal patterns

We find much larger differences between municipalities than between regions. In Figure 1C–D, we compare the ten most deprived municipalities with the ten least deprived. In Staro Nagorichane and Chashka, both small and rural, the headcount point estimates are 65.67 and 59.80 percent, respectively. In the next three municipalities, the point estimates place about half of households in poverty. At the other end, six municipalities, five of them urban municipalities of the capital, have point estimates that round to zero.

We find substantial variation in the width of the identification bounds across municipalities. Among the ten most deprived municipalities, Mogila has the widest range, from 17.45 to 90.36 percent, whereas Dojran has the narrowest, from 33.02 to 44.56 percent. The width depends on how tightly the geographic patterns and census margins jointly constrain the overlap of deprivations. Overall, the width of the bounds is strongly correlated with the point estimate (r = 0.76), so that uncertainty is largest precisely where poverty is highest.

Despite this uncertainty, we identify municipal comparisons that hold under any dependence structure. For instance, the lower bound for Staro Nagorichane (H = 31.13 percent, MPI = 0.1798) exceeds the upper bound for Aerodrom, a well-served urban municipality in the capital (H = 3.43 percent, MPI = 0.0126). Since each municipality's distribution can vary independently, the former is more deprived than the latter for every pair of distributions consistent with the data. Across all municipalities, 517 of the 6,320 ordered pairs are ranked in the same way under any dependence structure, and 35 municipalities are robustly more deprived than at least one other. This is a much stronger statement than a comparison of point estimates, and it shows that unlinked data can still deliver robust conclusions about where multidimensional poverty is concentrated.

At the same time, a point estimate of zero does not mean there is no poverty. In Aerodrom, only 3.1 percent of households are geographically deprived, all of them in kindergarten access, but all seven census indicators are positive. Under independence, these deprivations rarely coincide in the same household, so the point estimate rounds to zero. If they do coincide, however, up to 3.43 percent of households could be poor. More generally, 24 municipalities have a lower bound of zero. Hence, in well-served urban municipalities, where poverty depends almost entirely on the overlap of socio-economic deprivations, the independence assumption matters most.

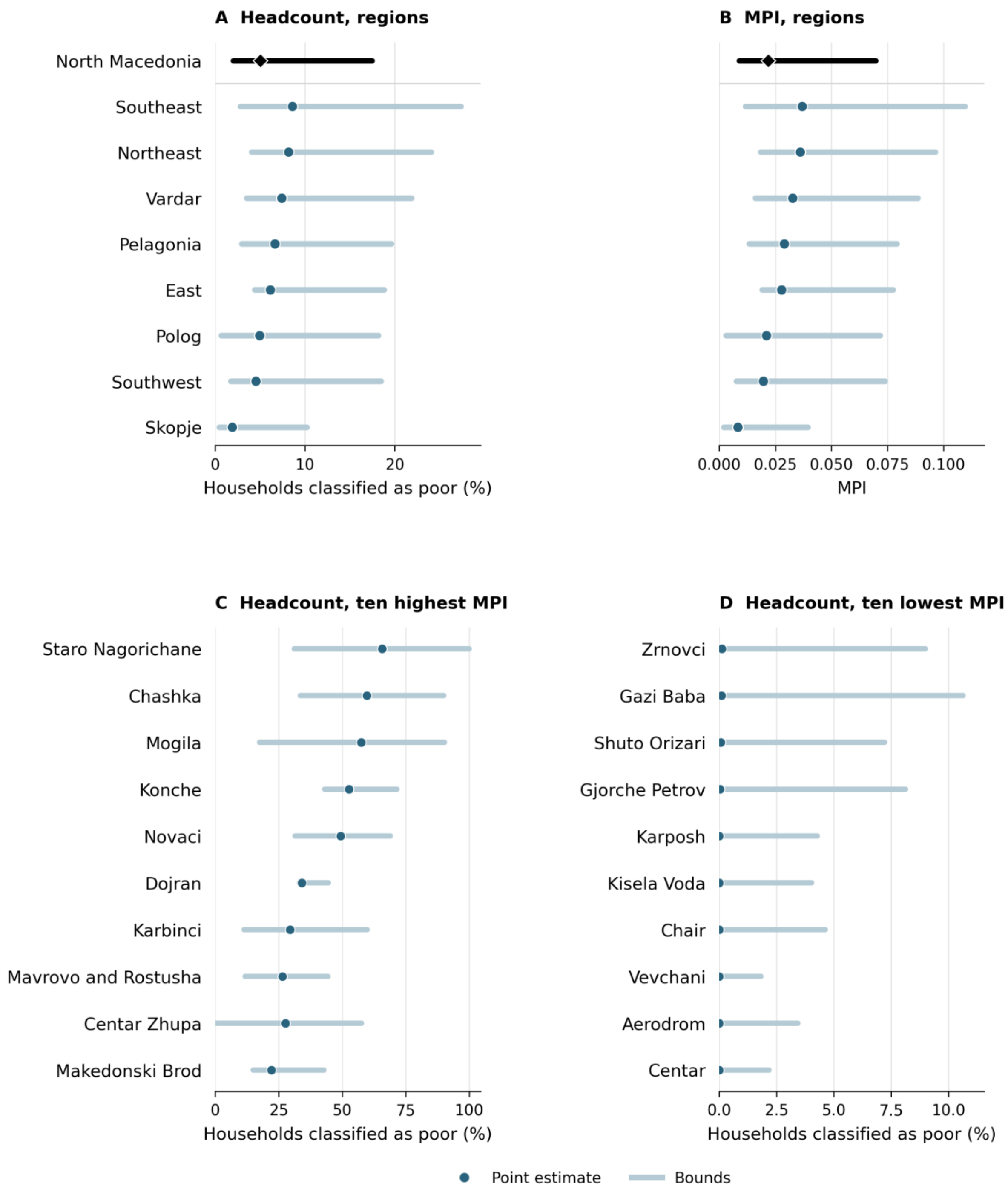

**Figure 1. Multidimensional poverty in regions and municipalities. A Headcount ratio and B MPI by statistical region; the national estimate is shown in black, and regions are ordered by MPI. C Headcount ratio in the ten municipalities with the highest MPI and D in the ten with the lowest, on different horizontal scales. Dots mark the independence point estimates and bars span the bounds, which are identification intervals, not confidence intervals.**

## 4.3 Which deprivations matter?

We next examine the national prevalence of each deprivation in Figure 2A. Unlike H and the MPI, these rates are directly observed and do not depend on any assumption about dependence. The rates vary by almost two orders of magnitude. Distance to a kindergarten is the most common deprivation, affecting 41.13 percent of households, followed by low adult education (26.10 percent) and distance to a secondary school (26.05 percent). At the other end, dilapidated housing (0.69 percent) and no heating (0.58 percent) are the rarest.

In Figure 2B, we split the average deprivation score into the contribution of each domain. Since the score is additive across indicators, its population mean (0.1106) decomposes exactly, again without any assumption about how deprivations overlap. Geographic access accounts for 69.0 percent of the average score and socio-economic conditions for 31.0 percent. The geographic domain dominates both because its indicators carry a higher individual weight (1/12 versus 1/14) and because kindergarten and secondary-school distance are among the most common deprivations. We note that this decomposition refers to the uncensored score. The contribution of each indicator to the MPI itself would require knowing how often it occurs among the poor, which our data do not identify.

We find a moderate correlation between the geographic and socio-economic scores across municipalities (r = 0.49), corresponding to about a quarter of their variance. Some of the most remote municipalities are not among the most socio-economically deprived. Combining the two datasets thus captures partly different patterns of deprivation.

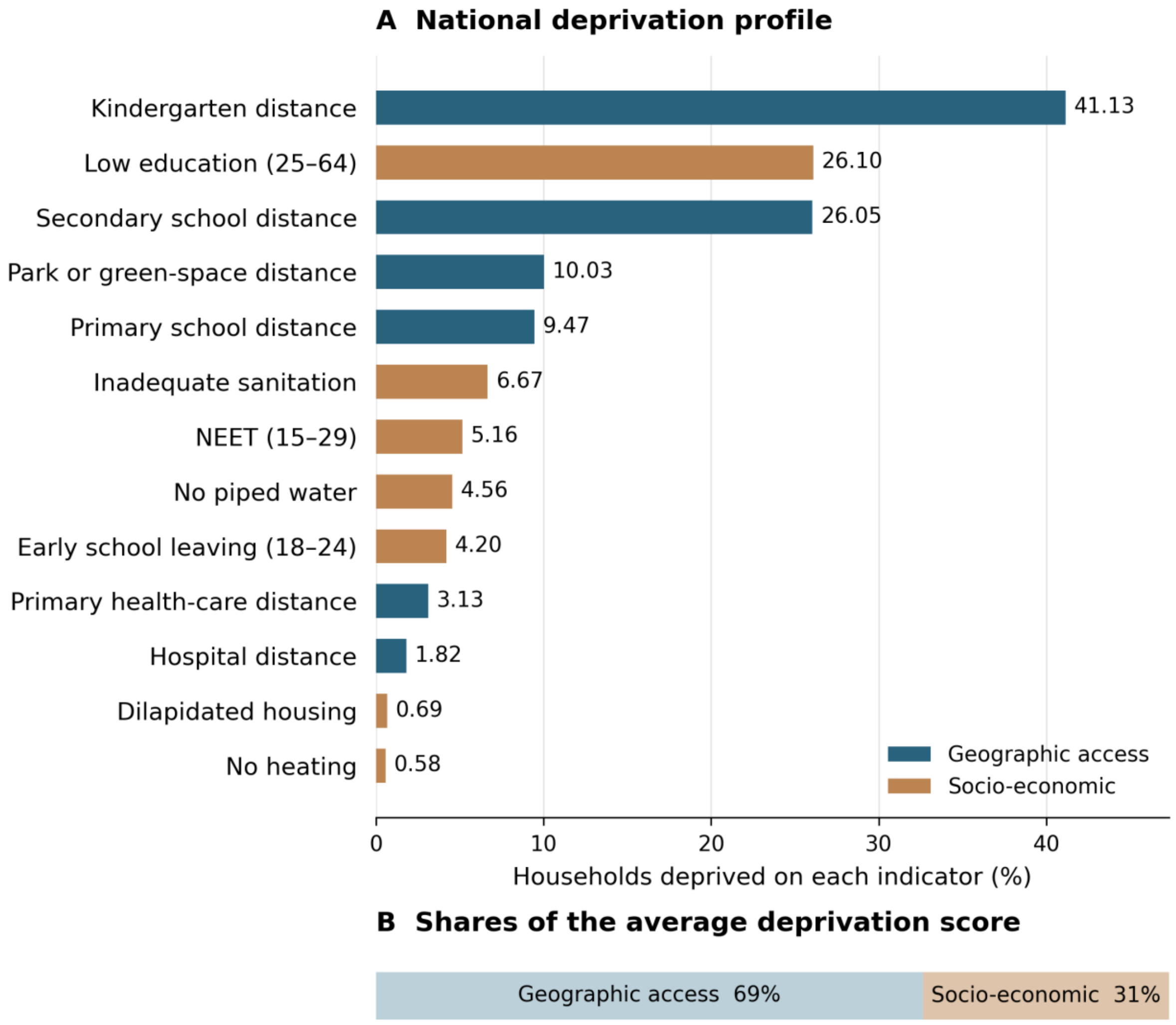

**Figure 2. Deprivation by indicator. A National deprivation rate of each indicator. Rates refer to all households, not only the poor; geographic rates assign each household the status of its settlement. B Share of the average deprivation score contributed by the geographic (blue) and socio-economic (ocher) domains.**

### 4.4 Robustness to measurement choices

Our index rests on three sets of choices: how much weight each domain receives, how far from a service a household must be to count as deprived, and which indicators enter the index. Since none of these choices is uniquely correct, we check whether our main conclusions survive reasonable alternatives. In each check, we change one choice, keep everything else as in the baseline, and recompute the point estimates and the bounds for all 80 municipalities.

We consider three types of alternatives. First, we change the domain weights. In the baseline, geographic access and socio-economic conditions each receive half of the weight. Instead, we give geographic access one third or two thirds of the weight, and, as a third alternative, give each of the thirteen indicators the same weight of 1/13. Second, we change the distance thresholds, multiplying all six of them by 0.5, 0.75, 1.25, and 1.5. For example, a household is deprived in primary-school access when the nearest school is more than 1.25 km away in the strictest case and more than 3.75 km away in the most lenient one, compared with 2.5 km in the baseline. Third, we drop each of the thirteen indicators in turn and divide its weight equally among the remaining indicators of its domain. Together, these checks give 20 alternative specifications.

For each alternative, we ask two questions. First, does the level of poverty move beyond the uncertainty that the unobserved overlap of deprivations already implies, that is, outside the bounds of the baseline? Second, does the ranking of municipalities change? We measure the latter by the Spearman correlation between the municipal MPI and the baseline. Table 3 summarizes the results by type of alternative, and Table B1 in Appendix B reports them for each alternative.

**Table 3. Robustness of the results to alternative measurement choices.**

| Type of alternative | Number of alternatives | National H % | Within bounds % | Spearman ρ |
|---|---|---|---|---|
| Baseline | – | 5.01 | – | 1.00 |
| Domain weights | 3 | 2.59–9.49 | 85 | 0.93 |
| Distance thresholds | 4 | 2.26–13.17 | 80 | 0.93 |
| One indicator dropped | 13 | 3.20–10.43 | 98 | 0.94 |

Note: Number of alternatives is how many alternative versions of the index we compute for each type: three sets of domain weights, four distance thresholds, and thirteen indicator sets, each without one indicator. National H is the range of the national headcount across the alternatives of each type; the baseline bounds are 2.01 to 17.44 percent. Within bounds is the share of all alternative municipal headcounts that lie within the baseline bounds of the same municipality. Spearman ρ is the lowest rank correlation between the municipal MPI and the baseline across the alternatives of each type. Results for each alternative are reported in Table B1.

The level of poverty varies with our choices, but within the range that our data already allow. The national headcount falls to 2.26 percent when all distance thresholds are 50 percent higher and rises to 13.17 percent when they are halved, and it ranges from 2.59 percent when geography receives a third of the weight to 9.49 percent when it receives two thirds. Dropping hospital or primary health-care access raises the headcount to about 10 percent, because their weight shifts to the more common access deprivations, whereas giving all thirteen indicators equal weight leaves it unchanged, since under both weighting schemes a household is poor if and only if it experiences at least five deprivations. Importantly, every one of these alternative headcounts lies within the bounds of the baseline (2.01 to 17.44 percent), and so do 92 percent of the alternative municipal estimates. Our measurement choices thus move the estimates less than the uncertainty about how deprivations overlap.

Moreover, the ranking of municipalities hardly changes. In every alternative, the Spearman correlation with the baseline is at least 0.93. In other words, our choices affect how much poverty we measure, but hardly where we find it. Table B1 also reports, for each alternative, how many of the ten most deprived municipalities remain among the ten most deprived, and how many municipal comparisons hold under any dependence structure.

Altogether, our results are robust to these measurement choices. The alternative levels of poverty lie within the bounds that our data already imply, and the ranking of municipalities remains essentially unchanged.

### 4.5 Validation against official statistics

Having measured multidimensional poverty, we next ask whether our estimates capture meaningful variation. Since the joint distribution of deprivations is unobserved, we cannot validate our estimates against household data directly. Instead, we compare them with three external sources: official statistics, the OPHI Global MPI, and a fieldwork-based deprivation index. None of them is expected to coincide with our index, since they differ in indicators, data, years, and populations. Rather, the comparisons show whether our index captures meaningful variation in poverty and which of its parts drives that variation.

We begin with official statistics published by the State Statistical Office. We select indicators that meet three criteria. First, they must be available at the local level, for all 80 municipalities or, failing that, for the eight regions. Second, they must play no role in the construction of our index. Other census statistics on education, employment, or housing would make the comparison circular, since they measure the same conditions as our socio-economic indicators. Third, they must be related to deprivation through clear channels. Four indicators meet these criteria: the share of the population aged 65 and over in 2021, the crude death and birth rates, computed as 2025 vital events per 1,000 people in the 2021 Census, and regional GDP per capita, computed as 2024 GDP divided by the 2021 population.

These indicators lead to different expectations for the two parts of our index. Population ageing and depopulation are concentrated in remote rural areas, so municipalities with an older population and a higher death rate should score higher on the geographic sub-index. For the socio-economic sub-index, we expect the opposite. Our education and NEET indicators can only be triggered by household members of working or school age, so households in older municipalities are less likely to be deprived in them, whereas younger and larger households, reflected in higher birth rates, are more likely to be. Finally, we expect both sub-indices to be lower in regions with a higher GDP per capita. For each indicator, we compute the correlation with the MPI and with each of its two sub-indices, where a sub-index is the average deprivation score within its domain. We plot the MPI against each of the four indicators in Figure 3A–D, show the same relationships separately for the two sub-indices in Figure 3E, and report all correlations in Table 4.

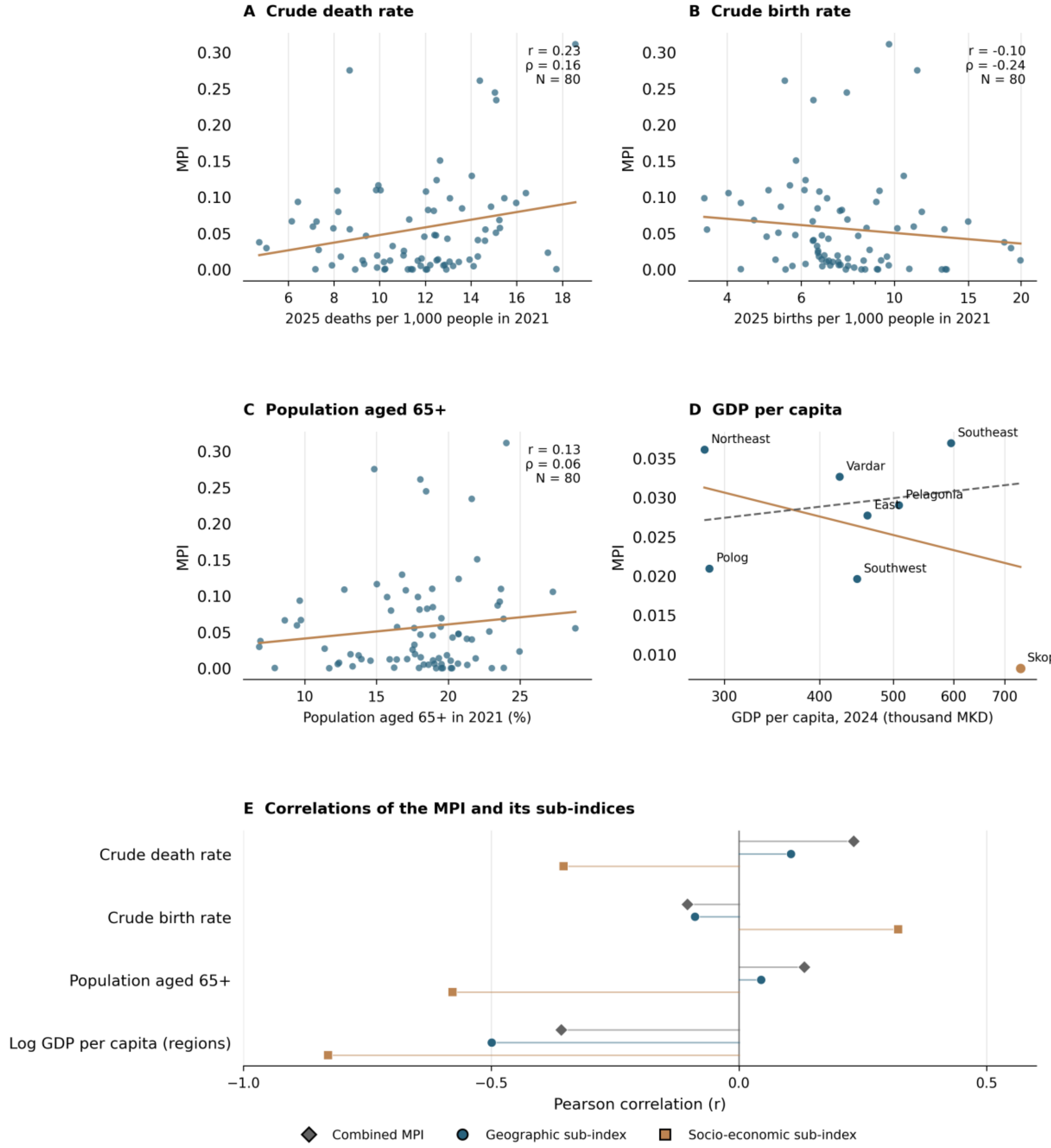

**Figure 3. MPI and official statistics. MPI against A the crude death rate, B the crude birth rate, and C the population share aged 65 and over, across 80 municipalities, and D against GDP per capita across eight regions. Lines show ordinary least-squares fits, with the horizontal axis in logs in B and D; in D, the dashed line excludes the capital region. E Pearson correlations of the combined MPI and its geographic and socio-economic sub-indices with the four indicators. Source: State Statistical Office and own calculations.**

**Table 4. Correlations between the index and official statistics.**

| External measure | N | Combined MPI r | Combined MPI rho | Geographic score r | Census score r |
|---|---|---|---|---|---|
| Deaths in 2025 per 1,000 people in 2021 | 80 | 0.2313 | 0.1610 | 0.1049 | -0.3538 |
| Births in 2025 per 1,000 people in 2021 | 80 | -0.1039 | -0.2391 | -0.0887 | 0.3210 |
| Population aged 65 and over in 2021 | 80 | 0.1314 | 0.0586 | 0.0447 | -0.5779 |
| Log of 2024 GDP divided by 2021 population | 8 | -0.3588 | -0.2143 | -0.4995 | -0.8294 |

Note: r is the Pearson and rho the Spearman rank correlation. Correlations are unweighted across municipalities (N = 80) or regions (N = 8). Geographic and census scores are the uncensored sub-indices.

We find only weak associations between the combined MPI and the three demographic indicators. It correlates weakly and positively with the death rate ($r = 0.23$, $\rho = 0.16$) and the share of the population aged 65 and over ($r = 0.13$), and weakly and negatively with the birth rate ($r = -0.10$). This is in line with our expectations: the two sub-indices respond to the age of the population in opposite directions, so their relationships largely offset each other in the combined index.

The sub-indices are in line with this. As expected, the socio-economic sub-index correlates negatively with the death rate ($r = -0.35$) and the share of older people ($r = -0.58$), and positively with the birth rate ($r = 0.32$). The geographic sub-index also has the expected signs, being higher where the death rate is higher ($r = 0.10$), the population older ($r = 0.04$), and the birth rate lower ($r = -0.09$), but these relationships are weak.

At the regional level, also as expected, multidimensional poverty is lower in richer regions. The MPI correlates negatively with GDP per capita ($r = -0.36$), and so do both sub-indices, the socio-economic one strongly ($r = -0.83$) and the geographic one moderately ($r = -0.50$). For the combined MPI, this relationship is driven by the capital region, which is both by far the richest region and the least deprived. Excluding it, the remaining seven regions show a weak positive relationship ($r = 0.20$), although seven observations are too few to distinguish a weak relationship from none.

In sum, the socio-economic sub-index behaves as expected across all four indicators, whereas the geographic sub-index does so only weakly. This is because the geographic sub-index measures physical remoteness from services, a dimension of deprivation that these demographic and economic statistics do not capture directly.

### 4.6 Validation against survey data

Next, we compare our index with the only other multidimensional poverty measure available for the country's regions, the OPHI Global MPI. It is based on the 2018–2019 Multiple Indicator Cluster Survey and, like our index, uses a cutoff of one third, but it combines ten different indicators of acute poverty, measured for individuals rather than households (Alkire and Santos, 2014). At the national level, OPHI reports a headcount of 0.4 percent and an MPI of 0.001 (OPHI, 2024). Our estimates are considerably higher because our index includes access to services and uses different education, employment, and housing indicators. The difference in levels is therefore not a correction of OPHI's estimate, and we compare only the regional rankings.

We compare the two rankings in Figure 4A. Overall agreement is weak (Spearman $\rho = 0.08$). Both indices agree on the most deprived region. The largest divergence is the capital region, which our index ranks as the least deprived and OPHI as the second most deprived. Excluding it, agreement across the remaining seven regions becomes moderate ($\rho = 0.38$). Neither sub-index, taken alone, agrees with the OPHI ranking ($\rho = -0.24$ for the geographic and $-0.30$ for the socio-economic sub-index), which suggests that the two measures capture different combinations of deprivations.

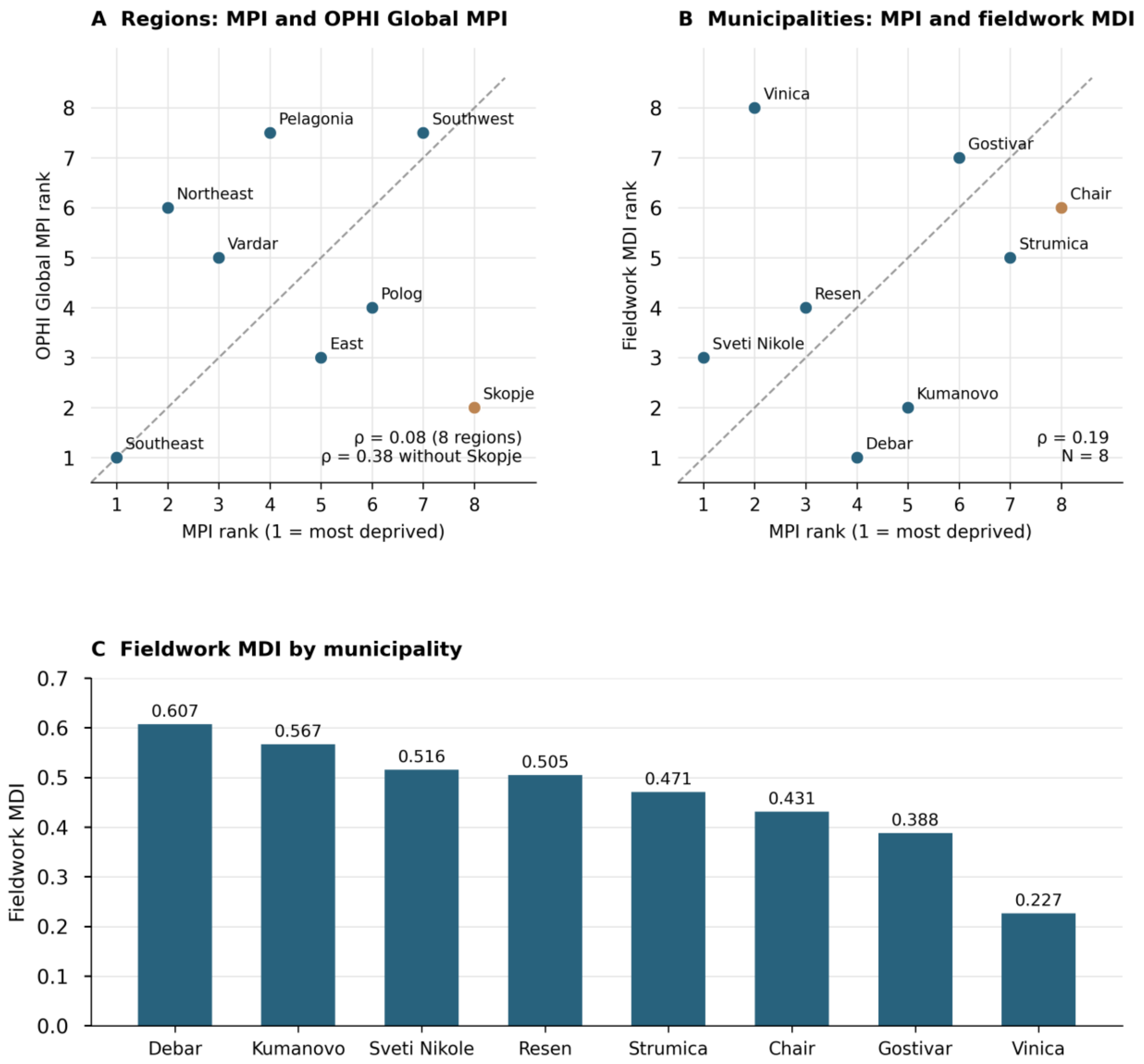

**Figure 4. Comparison with the OPHI Global MPI and the fieldwork index. A Regional ranks under our MPI and the OPHI Global MPI; tied OPHI ranks receive average ranks. B Municipal ranks under our MPI and the fieldwork MDI in the eight matched municipalities. In A and B, rank 1 is the most deprived and the diagonal marks perfect agreement. C Fieldwork MDI in each matched municipality. Source: own calculations, OPHI (2024), and Gerovska Mitev et al. (2026).**

## 4.7 Validation against the fieldwork index

Finally, we compare our index with the Multidimensional Deprivation Index (MDI) of a companion study (Gerovska Mitev et al., 2026), the highest-resolution comparison available to us. The study surveyed 152 social welfare beneficiaries in eight municipalities, one in each statistical region, and combines eight indicators of education, employment, welfare dependence, self-assessed hardship, household structure, and disability. We compare the eight matched municipalities in Table 5 and Figure 4B–C. As expected, deprivation among welfare beneficiaries is much higher than in the general population, with fieldwork headcounts between 50 and 100 percent compared with between 0 and 11 percent in our index. For this reason, we focus on the rankings.

We find weak rank agreement with the combined MPI (Spearman ρ = 0.19) and little linear association between the two measures across the eight municipalities (Pearson r = −0.17). Split into the two sub-indices, however, our socio-economic ranking tracks the fieldwork MDI considerably more closely (ρ = 0.33, r = 0.43) than the geographic one (ρ = 0.19, r = 0.02). This is consistent with the fieldwork index's focus on education, employment, and material hardship, and with the pattern we observed against official statistics. Rank agreement with the fieldwork measure is higher for the upper-bound headcount map (ρ = 0.52) than for the point-estimate headcount map (ρ = 0.29). This comparison concerns the ordering of municipalities and does not establish whether true headcounts are closer to the upper bounds.

**Table 5. Our MPI and the fieldwork MDI in the eight matched municipalities.**

| Municipality | Spatial H % | Spatial MPI | Beneficiary H % | Beneficiary MDI |
|---|---|---|---|---|
| Chair | 0.00 | 0.0000 | 72.2 | 0.431 |
| Debar | 4.34 | 0.0177 | 97.1 | 0.607 |
| Gostivar | 2.38 | 0.0103 | 70.0 | 0.388 |
| Kumanovo | 2.80 | 0.0122 | 100.0 | 0.567 |
| Resen | 9.11 | 0.0397 | 95.7 | 0.505 |
| Strumica | 1.16 | 0.0050 | 92.3 | 0.471 |
| Sveti Nikole | 10.65 | 0.0471 | 95.7 | 0.516 |
| Vinica | 8.90 | 0.0406 | 50.0 | 0.227 |

Note: Spatial estimates refer to all households under the independence point estimate. Beneficiary estimates refer to the purposive fieldwork sample of the companion study (Gerovska Mitev et al., 2026). Levels are not comparable across the two measures.

We illustrate this divergence with Chair, an urban municipality in the capital. Our point estimate rounds to zero, ranking it 77th of 80 municipalities, yet the fieldwork classifies 72.2 percent of the beneficiaries interviewed there as deprived. None of its households is geographically deprived in any indicator, so its poverty depends entirely on whether socio-economic deprivations coincide in the same households, which is exactly the information our data lack. This example underlines why a low municipal point estimate should be reported with its bounds. The fieldwork sample consists of welfare beneficiaries and uses different indicators, so the two rates cannot be treated as estimates for the same population.

## 5 Discussion

Local policy needs local poverty estimates, yet the data needed to construct a multidimensional poverty index are rarely available below the national or regional level. In this paper, we showed how to measure multidimensional poverty at the local level when the relevant data cannot be linked at the household level. Using North Macedonia as a case study, we combined settlement-level access to services with municipal census data and, following Brunckhorst et al. (2026), paired an independence-based point estimate with sharp bounds for all 80 municipalities.

Our results show that incomplete data can still reveal a great deal about where multidimensional poverty is concentrated. The national bounds, from 2.01 to 17.44 percent, are wide, and the regional ranking cannot be established without assumptions about how deprivations overlap. At the municipal level, however, 517 comparisons hold under every joint distribution consistent with the data, and the most deprived municipalities, all small and rural, are poorer than nearly all municipalities of the capital. Moreover, the ranking of municipalities hardly changed when we varied the weights, distance thresholds, and indicators, and the alternative poverty levels remained within the baseline bounds. Hence, the main uncertainty in our estimates concerns the level of poverty, while its location is much more certain.

Our comparisons with external sources also clarify what the two parts of the index capture. The socio-economic sub-index behaved as expected against official statistics and was more closely aligned with the fieldwork index, whereas the geographic sub-index was only weakly related to these measures, and the combined MPI agreed only weakly with the OPHI regional ranking. However, access to services remains relevant. Remoteness is a distinct dimension of deprivation that existing statistics do not capture, and the weight it should receive in a local MPI is a normative choice that depends on the purpose of the measure.

These findings have two implications for policy. First, the municipal estimates can guide where to act. The five municipalities in which more than 30 percent of households are poor even under the most favorable dependence structure warrant attention under any assumption, and the settlement-level access indicators identify the places that are far from kindergartens, schools, and health facilities. Second, the estimates cannot replace household-level assessment. They identify poor places, while assistance to individual households still requires information on the households themselves. The case of Chair, with a point estimate close to zero and most of the interviewed beneficiaries deprived, shows this clearly.

A multidimensional tool also shows who is poor and why. Income poverty and multidimensional poverty overlap only partly, and groups such as older people and households with members with disabilities face different risks, which is important for decisions on whom to target (Peng et al., 2024). The geographic and the socio-economic parts of our index also capture different dimensions, which is why both belong in one index. Hence, the index may be used as a planning tool in different sectors (Mazenda and Sari, 2026). Moreover, an index compiled from administrative and geographic data moves the burden of proof away from the people it aims to reach (Herd and Moynihan, 2018). However, it should not be used to label households or locations, as labeling may lead households to leave programs out of shame (Habibullah et al., 2023).

The fieldwork in North Macedonia shows what such an instrument would capture (Gerovska Mitev et al., 2026). Deprivation among social assistance beneficiaries includes loss of dignity, fear of losing benefits and social exclusion, as well as irregular work, poor housing and unmet health needs. Professionals stress education and individual human capital as causes of poverty, while beneficiaries point to the lack of jobs, discrimination and institutional failures (Gerovska Mitev et al., 2026). Therefore, the tool can complement the income-based approach and should be part of an integrated system of records linking centres for social work with employment services, health and educational institutions.

Our results also point to a simple and inexpensive improvement in data provision. Statistical offices could publish census cross-tabulations that report how pairs of deprivations overlap, alongside the totals for each indicator. Even a few such tables would add constraints to the feasible joint distributions and could narrow the bounds considerably.

Nevertheless, our study is not without its limitations, some of which we addressed in this paper and others that require further work.

First, and most importantly, our point estimate assumes that the census deprivations are independent of each other and of access to services within each municipality. Yet deprivations tend to cluster within households (Suppa et al., 2022; Suppa, 2025), and the relationships between them are central to integrated policies that target the deprivations whose improvement spills over to others (Stojkoski et al., 2025). We mitigated this limitation in two ways. We preserved the joint distribution of the geographic deprivations, which we observe directly. We reported sharp bounds and based our main conclusions on the comparisons that hold under any dependence structure. Still, the bounds do not reveal the actual relationships, so our estimates can inform where to act, but not which combination of deprivations to target within a municipality. Future work could address this by constraining the joint distribution with empirical information on co-deprivation, such as census cross-tabulations or the Poverty Space estimated from household surveys (Stojkoski et al., 2025), for the indicators that both sources share, such as sanitation, housing, water, and education.

Second, the index rests on choices about indicators, distance thresholds, weights, and the poverty cutoff, all of which can affect multidimensional poverty measures (Decancq and Lugo, 2013; Azpitarte et al., 2020). We anchored the distance thresholds in national norms wherever they exist, followed the standard nested weighting, and tested 20 alternative specifications. The municipal ranking was robust to all of them, although the poverty cutoff, including whether households exactly at it count as poor, shifts the level of poverty beyond the baseline bounds (Appendix A). Moreover, our indicators are tailored to North Macedonia, which makes them locally relevant but harder to compare internationally (Santos, 2019). Future applications could derive the weights from the preferences of experts and policymakers (Carrino et al., 2024), and include a subset of internationally comparable indicators.

Third, the geographic indicators measure road distance from a single point in each settlement, and do not measure travel time, transport, or service quality (Geurs and van Wee, 2004). They apply to every household, including households without children, and refer to a different year than the census data. We limited these concerns by measuring access separately for each of the 1,575 settlements, and by varying all distance thresholds in our robustness checks. Data on public transport, facility capacity, and household composition would allow future work to measure access as travel time to services of adequate quality, for the households that actually need them.

Fourth, our bounds capture uncertainty about how deprivations overlap, but not measurement error in the data, and are not confidence intervals. We cross-checked settlement and facility locations against several sources. Combining the bounds with resampling of the inputs would allow future work to account for both sources of uncertainty.

Finally, our validation exercises rely on few observations, especially for the eight regions, and the fieldwork index covers a purposive sample of welfare beneficiaries. We partly offset this by comparing our index with three external sources, which differ in their data, indicators, and populations. New rounds of household surveys, and fieldwork in more municipalities, would allow a more direct validation of the municipal estimates.

While imperfect, our approach offers a practical way to measure multidimensional poverty at the local level in countries where linked household data are unavailable. As linked data and empirical evidence on co-deprivation become available, the same framework can incorporate them, narrowing the bounds and allowing local estimates to inform both where to act and which deprivations to target.

## Acknowledgments

This paper was prepared within the research project „Детерминанти на повеќедимензионалната сиромаштија: просторно мапирање и нови одредници за социјалните и за развојните политики на национално ниво“ [Determinants of multidimensional poverty: Spatial mapping and new guidelines for social and development policies at the national level], funded by the Ministry of Education and Science of the Republic of North Macedonia. An interactive dashboard presenting the results is available at the Multidimensional Poverty Atlas (https://mdpovertyatlas.mk).

## Data availability

The analysis uses census tabulations, settlement household totals, road-network data, and facility registers. The code, processed inputs, and outputs needed to reproduce every table and figure are provided in the accompanying reproducibility package. Road-network data are © OpenStreetMap contributors, available under the Open Database License. The package is available at https://doi.org/10.5281/zenodo.24108030.

## References

Alkire, S., and Ballon, P. (2012). Understanding association across deprivation indicators in multidimensional poverty. Research Workshop on Dynamic Comparison between Multidimensional Poverty and Monetary Poverty. OPHI, University of Oxford.

Alkire, S., and Foster, J. (2011a). Counting and multidimensional poverty measurement. Journal of Public Economics, 95(7–8), 476–487. https://doi.org/10.1016/j.jpubeco.2010.11.006

Alkire, S., and Foster, J. (2011b). Understandings and misunderstandings of multidimensional poverty measurement. The Journal of Economic Inequality, 9, 289–314. https://doi.org/10.1007/s10888-011-9181-4

Alkire, S., and Santos, M. E. (2014). Measuring acute poverty in the developing world: Robustness and scope of the Multidimensional Poverty Index. World Development, 59, 251–274. https://doi.org/10.1016/j.worlddev.2014.01.026

Alkire, S., and Seth, S. (2015). Multidimensional poverty reduction in India between 1999 and 2006: Where and how? World Development, 72, 93–108.

Alkire, S., Foster, J., Seth, S., Santos, M. E., Roche, J. M., and Ballon, P. (2015). Multidimensional Poverty Measurement and Analysis. Oxford University Press. https://doi.org/10.1093/acprof:oso/9780199689491.001.0001

Alkire, S., Kanagaratnam, U., and Suppa, N. (2024). The global Multidimensional Poverty Index (MPI) 2024 disaggregation results and methodological note. OPHI MPI Methodological Note 59. Oxford Poverty and Human Development Initiative, University of Oxford. https://ophi.org.uk/publications/MN-59

Anand, P., Jones, S., Donoghue, M., and Teitler, J. (2021). Non-monetary poverty and deprivation: A capability approach. Journal of European Social Policy, 31(1), 78–91. https://doi.org/10.1177/0958928720938334

Anselin, L. (1995). Local indicators of spatial association—LISA. Geographical Analysis, 27(2), 93–115. https://doi.org/10.1111/j.1538-4632.1995.tb00338.x

Atkinson, A. B. (2003). Multidimensional deprivation: Contrasting social welfare and counting approaches. The Journal of Economic Inequality, 1, 51–65. https://doi.org/10.1023/A:1023903525276

Azpitarte, F., Gallegos, J., and Yalonetzky, G. (2020). On the robustness of multidimensional counting poverty orderings. The Journal of Economic Inequality, 18, 339–364. https://doi.org/10.1007/s10888-019-09435-5

Benjamini, Y., and Hochberg, Y. (1995). Controlling the false discovery rate: A practical and powerful approach to multiple testing. Journal of the Royal Statistical Society Series B, 57(1), 289–300.

Bossert, W., Chakravarty, S. R., and D'Ambrosio, C. (2013). Multidimensional poverty and material deprivation with discrete data. Review of Income and Wealth, 59(1), 29–43.

Bourguignon, F., and Chakravarty, S. R. (2003). The measurement of multidimensional poverty. The Journal of Economic Inequality, 1, 25–49. https://doi.org/10.1023/A:1023913831342

Brunckhorst, B. J., Nguyen, M. C., Yonzan, N., Dang, H.-A., and Lakner, C. (2026). Fusing data sources to measure multidimensional poverty. Policy Research Working Paper 11424. World Bank. https://documents.worldbank.org/curated/en/099536107012637783/pdf/IDU-63ce812c-0a33-4c62-bfb9-5f162b6ef539.pdf

Carrino, L., Farnia, L., and Giove, S. (2024). Measuring social inclusion in Europe: A non-additive approach with the expert-preferences of public policy planners. Journal of the Royal Statistical Society Series A: Statistics in Society, 187(1), 231–259. https://doi.org/10.1093/jrsssa/qnad106

CONEVAL. (2021). Metodología para la medición de la pobreza en los municipios de México, 2020. Consejo Nacional de Evaluación de la Política de Desarrollo Social, Mexico. https://www.coneval.org.mx/Medicion/Documents/Pobreza_municipal/2020/Metodologia_pobreza_municipal_2020.pdf

Corral, P., Henderson, H., and Segovia, S. (2025). Poverty mapping in the age of machine learning. Journal of Development Economics, 172, 103377. https://doi.org/10.1016/j.jdeveco.2024.103377

DANE. (2020). Medida de pobreza multidimensional municipal de fuente censal: Boletín técnico, Censo Nacional de Población y Vivienda 2018. Departamento Administrativo Nacional de Estadística, Bogotá. https://www.dane.gov.co/files/investigaciones/condiciones_vida/pobreza/2018/informacion-censal/bt-censal-pobreza-municipal-2018.pdf

de Neubourg, C., Chai, J., de Milliano, M., and Plavgo, I. (2013). Step-by-step guidelines to the Multiple Overlapping Deprivation Analysis (MODA). Innocenti Working Papers. UNICEF Office of Research, Florence.

Decancq, K. (2014). Copula-based measurement of dependence between dimensions of well-being. Oxford Economic Papers, 66(3), 681–701. https://doi.org/10.1093/oep/gpt038

Decancq, K., and Lugo, M. A. (2013). Weights in multidimensional indices of wellbeing: An overview. Econometric Reviews, 32(1), 7–34. https://doi.org/10.1080/07474938.2012.690641

Diaz-Bonilla, C., Sabatino, C., Wu, H., and Nguyen, M. C. (2022). October 2022 update to the Multidimensional Poverty Measure: What's new. Global Poverty Monitoring Technical Note 26. World Bank.

D'Orazio, M., Di Zio, M., and Scanu, M. (2006). Statistical matching for categorical data: Displaying uncertainty and using logical constraints. Journal of Official Statistics, 22(1), 137–157. https://www.istat.it/wp-content/uploads/2014/04/jos-2006-221.pdf

Duclos, J.-Y., Sahn, D. E., and Younger, S. D. (2006). Robust multidimensional poverty comparisons. The Economic Journal, 116(514), 943–968. https://doi.org/10.1111/j.1468-0297.2006.01118.x

Elbers, C., Lanjouw, J. O., and Lanjouw, P. (2003). Micro-level estimation of poverty and inequality. Econometrica, 71(1), 355–364. https://doi.org/10.1111/1468-0262.00399

Feres, J. C., and Mancero, X. (2001). El método de las necesidades básicas insatisfechas (NBI) y sus aplicaciones en América Latina. Estudios Estadísticos y Prospectivos, LC/L.1491-P. Economic Commission for Latin America and the Caribbean (ECLAC), Santiago. https://repositorio.cepal.org/handle/11362/4784

Ferreira, F. H. G., and Lugo, M. A. (2013). Multidimensional poverty analysis: Looking for a middle ground. The World Bank Research Observer, 28(2), 220–235. https://doi.org/10.1093/wbro/lks013

Gerovska Mitev, M., Trbojevikj, S., Miovska-Spaseva, S., Jovcheska, S., and Stojkoski, V. (2026). Повеќедимензионална сиромаштија: перцепции и состојби на локално ниво [Multidimensional poverty: Perceptions and conditions at the local level]. Skopje: Faculty of Philosophy, Ss. Cyril and Methodius University in Skopje.

Geurs, K. T., and van Wee, B. (2004). Accessibility evaluation of land-use and transport strategies: Review and research directions. Journal of Transport Geography, 12(2), 127–140. https://doi.org/10.1016/j.jtrangeo.2003.10.005

Habibullah, H., Yuda, T. K., Setiawan, H. H., and Susantyo, B. (2023). Moving beyond stereotype: A qualitative study of long-standing recipients of the Indonesian conditional cash transfers (CCT/PKH). Social Policy & Administration, 58(1), 108–121. https://doi.org/10.1111/spol.12946

Herd, P., and Moynihan, D. P. (2018). Administrative Burden: Policymaking by Other Means. New York: Russell Sage Foundation.

Huangfu, Q., and Hall, J. A. J. (2018). Parallelizing the dual revised simplex method. Mathematical Programming Computation, 10(1), 119–142.

Jean, N., Burke, M., Xie, M., Davis, W. M., Lobell, D. B., and Ermon, S. (2016). Combining satellite imagery and machine learning to predict poverty. Science, 353(6301), 790–794. https://doi.org/10.1126/science.aaf7894

Kovesdi, F., and Alkire, S. (2026). Measuring multidimensional poverty: A rigorously documented global data gap. Social Indicators Research, 182, article 2. https://doi.org/10.1007/s11205-025-03801-9

Manski, C. F. (2003). Partial Identification of Probability Distributions. Springer. https://doi.org/10.1007/b97478

Mazenda, A., and Sari, N. P. W. P. (2026). Multidimensional poverty and welfare in developing urban areas. Discover Public Health, 23, 715. https://doi.org/10.1186/s12982-026-02045-3

McLennan, D., Noble, S., Noble, M., Plunkett, E., Wright, G., and Gutacker, N. (2019). The English Indices of Deprivation 2019: Technical report. Ministry of Housing, Communities and Local Government, London.

Menshikov, V., Kokina, I., Komarova, V., Ruza, O., and Danilevica, A. (2020). Measuring multidimensional poverty within the resource-based approach: A case study of Latgale region, Latvia. Entrepreneurship and Sustainability Issues, 8(2), 1211–1227. https://doi.org/10.9770/jesi.2020.8.2(72)

Molina, I., and Rao, J. N. K. (2010). Small area estimation of poverty indicators. The Canadian Journal of Statistics, 38(3), 369–385. https://doi.org/10.1002/cjs.10051

Moran, P. A. P. (1950). Notes on continuous stochastic phenomena. Biometrika, 37(1–2), 17–23. https://doi.org/10.2307/2332142

Nájera Catalán, H. E. (2018). Modelling severity of child deprivation, access to green spaces and availability of infrastructure in Mexico City. Paper presented at the International Workshop "Addressing Inclusion among Children and Adolescents Living in Poverty: Progress towards Achieving the SDGs", The New School and UNDP Headquarters, New York.

Nolan, B., and Whelan, C. T. (2010). Using non-monetary deprivation indicators to analyze poverty and social exclusion: Lessons from Europe? Journal of Policy Analysis and Management, 29(2), 305–325. https://doi.org/10.1002/pam.20493

Oxford Poverty and Human Development Initiative (OPHI). (2024). Global MPI Country Briefing 2024: North Macedonia. University of Oxford. https://ophi.org.uk/media/46035/download

Peng, C., Chang, Q., Wang, J. S.-H., Yeung, C. Y., and Yip, P. S. F. (2024). Patterns and determinants of multidimensional poverty and welfare interventions: Towards evidence-based poverty-alleviation policies in Hong Kong. International Journal of Social Welfare, 33(4), 931–950. https://doi.org/10.1111/ijsw.12648

Ravallion, M. (2011). On multidimensional indices of poverty. The Journal of Economic Inequality, 9, 235–248. https://doi.org/10.1007/s10888-011-9173-4

Ridder, G., and Moffitt, R. (2007). The econometrics of data combination. In J. J. Heckman and E. E. Leamer (Eds.), Handbook of Econometrics, Vol. 6B, pp. 5469–5547. Elsevier. https://doi.org/10.1016/S1573-4412(07)06075-8

Rippin, N. (2010). Poverty severity in a multidimensional framework: The issue of inequality between dimensions. Courant Research Centre: Poverty, Equity and Growth Discussion Paper No. 47. University of Göttingen.

Santos, M. E. (2019). Challenges in designing national multidimensional poverty measures. Statistics Series No. 100 (LC/TS.2019/5). Economic Commission for Latin America and the Caribbean (ECLAC). https://repositorio.cepal.org/handle/11362/44453

Sen, A. (1985). Commodities and Capabilities. North-Holland.

Sowula, J., Gehrig, F., Scruggs, L. A., Seeleib-Kaiser, M., and Ramalho Tafoya, G. (2024). The end of welfare states as we know them? A multidimensional perspective. Social Policy & Administration, 58(5), 785–799. https://doi.org/10.1111/spol.12990

Srbinoski, B., Petreski, B., and Petreski, M. (2023). Measurement of multidimensional child poverty: Evidence from North Macedonia. Child Indicators Research, 16(1), 247–271. https://doi.org/10.1007/s12187-022-09967-9

Stojkoski, V., Lopez-Calva, L. F., Bolch, K., and Fernandez, A. (2025). "Development acupuncture:" Mapping the network structure of multidimensional poverty. Journal of Economic Policy Reform, 28(4), 393–412. https://doi.org/10.1080/17487870.2025.2529195

Suppa, N. (2018). Towards a multidimensional poverty index for Germany. Empirica, 45(4), 655–683. https://doi.org/10.1007/s10663-017-9385-3

Suppa, N. (2025). Deprivations rarely come alone: Multidimensional poverty dynamics in Europe. Review of Income and Wealth, 71(4).

Suppa, N., Alkire, S., and Nogales, R. (2022). The many forms of poverty: Analyses of deprivation interlinkages in the developing world. OPHI Research in Progress 63a. Oxford Poverty and Human Development Initiative, University of Oxford.

Tarozzi, A., and Deaton, A. (2009). Using census and survey data to estimate poverty and inequality for small areas. The Review of Economics and Statistics, 91(4), 773–792. https://doi.org/10.1162/rest.91.4.773

Tsui, K.-Y. (2002). Multidimensional poverty indices. Social Choice and Welfare, 19, 69–93. https://doi.org/10.1007/s355-002-8326-3

Tuna, E., Simonovska, A., and Petrovska Mitrevska, B. (2022). Multidimensional analysis of rural poverty in North Macedonia. Journal of Agricultural, Food and Environmental Sciences, 76(8), 52–61. https://doi.org/10.55302/JAFES22768052e

Uganda Bureau of Statistics and UNICEF. (2019). The Geography of Multidimensional Poverty in Uganda. Kampala: UNICEF Uganda.

Weiss, D. J., Nelson, A., Gibson, H. S., et al. (2018). A global map of travel time to cities to assess inequalities in accessibility in 2015. Nature, 553, 333–336. https://doi.org/10.1038/nature25181

Whelan, C. T., Nolan, B., and Maître, B. (2014). Multidimensional poverty measurement in Europe: An application of the adjusted headcount approach. Journal of European Social Policy, 24(2), 183–197. https://doi.org/10.1177/0958928713517914

World Bank. (2018). Poverty and Shared Prosperity 2018: Piecing Together the Poverty Puzzle. World Bank, Washington, DC. https://doi.org/10.1596/978-1-4648-1330-6

## Appendix A Cutoff sensitivity

In Table A1, we report the national headcount under alternative poverty cutoffs, holding indicators, weights, and data constant. As expected, the headcount falls as the cutoff rises, from 21.15 percent at a cutoff of 0.20 to 0.37 percent at 0.50. The bounds remain wide relative to the point estimate at every cutoff, and at 0.50 the lower bound reaches zero while the upper bound is 7.76 percent. Both the cutoff and the unknown overlap of deprivations affect the headcount estimates. Moreover, since a household deprived in exactly four geographic indicators scores exactly one third, the baseline results also depend on whether the cutoff is strict. Counting such households as poor raises the point estimate of the headcount from 5.01 to 6.77 percent, and of the MPI from 0.0219 to 0.0277.

**Table A1. National headcount ratio under alternative poverty cutoffs.**

| Strict score cutoff | H estimate % | H lower bound % | H upper bound % |
|---|---|---|---|
| c > 0.20 | 21.15 | 14.05 | 37.31 |
| c > 1/3 | 5.01 | 2.01 | 17.44 |
| c > 0.40 | 4.08 | 2.01 | 13.28 |
| c > 0.50 | 0.37 | 0.00 | 7.76 |

Note: Households are poor when their score strictly exceeds the cutoff. Zeros are rounded values.

## Appendix B Robustness to measurement choices

In Table B1, we report the national headcount and its bounds under alternative domain weights, distance thresholds, and indicator sets, together with three measures of how the municipal pattern compares with the baseline: the Spearman correlation of the municipal MPI, the number of the ten most deprived municipalities that remain among the ten most deprived, and the number of ordered pairs of municipalities ranked in the same way under any dependence structure. When an indicator is excluded, the weight of its domain is divided equally among the remaining indicators of that domain. Distance thresholds are scaled jointly for all six access indicators.

**Table B1. Robustness of the results to measurement choices.**

| Specification | H % | H bounds % | Spearman ρ | Top 10 | Robust pairs |
|---|---|---|---|---|---|
| Baseline | 5.01 | 2.01–17.44 | 1.00 | 10 | 517 |
| Geographic domain weight 1/3 | 2.59 | 0.02–15.89 | 0.93 | 6 | 3 |
| Geographic domain weight 2/3 | 9.49 | 6.31–19.68 | 0.98 | 6 | 1,560 |
| Equal indicator weights (1/13) | 5.01 | 2.01–17.44 | 1.00 | 9 | 517 |
| Distance thresholds × 0.5 | 13.17 | 8.01–27.48 | 0.94 | 6 | 1,356 |
| Distance thresholds × 0.75 | 7.89 | 3.78–21.43 | 0.98 | 8 | 874 |
| Distance thresholds × 1.25 | 3.39 | 1.17–15.15 | 0.95 | 8 | 337 |
| Distance thresholds × 1.5 | 2.26 | 0.70–13.42 | 0.93 | 7 | 238 |
| Without kindergarten access | 4.92 | 2.01–15.69 | 1.00 | 8 | 624 |
| Without primary school access | 8.06 | 3.64–21.48 | 0.95 | 6 | 784 |
| Without secondary school access | 5.26 | 2.02–16.98 | 0.99 | 9 | 596 |
| Without primary health-care access | 10.28 | 5.23–24.15 | 0.96 | 6 | 1,118 |
| Without hospital access | 10.43 | 5.76–24.06 | 0.96 | 6 | 1,167 |
| Without green-space access | 7.85 | 2.66–21.82 | 0.95 | 7 | 560 |
| Without low education | 3.20 | 2.01–12.00 | 0.94 | 8 | 846 |
| Without early school leaving | 4.65 | 2.01–15.39 | 1.00 | 9 | 671 |
| Without NEET | 4.67 | 2.01–15.17 | 1.00 | 10 | 674 |
| Without inadequate sanitation | 4.49 | 2.01–14.65 | 0.99 | 9 | 698 |
| Without dilapidated housing | 4.95 | 2.01–17.13 | 1.00 | 9 | 534 |
| Without heating | 4.98 | 2.01–17.15 | 1.00 | 9 | 538 |
| Without piped water | 4.57 | 2.01–15.48 | 0.99 | 9 | 636 |

Note: H is the national headcount ratio. Spearman ρ compares the municipal MPI with the baseline. Top 10 is the number of the ten most deprived municipalities in the baseline that remain among the ten most deprived. Robust pairs is the number of the 6,320 ordered pairs of municipalities in which the lower bound of one exceeds the upper bound of the other.

## Appendix C Exploratory spatial association

Figure C1 maps the municipal point estimates of the MPI (A), the width of the headcount bounds (B), the number of households in each settlement (C), and the number of access deprivations in each settlement (D). The poorest municipalities are scattered across the country rather than concentrated in particular regions, and settlements far from services lie next to well-served ones, often within the same municipality.

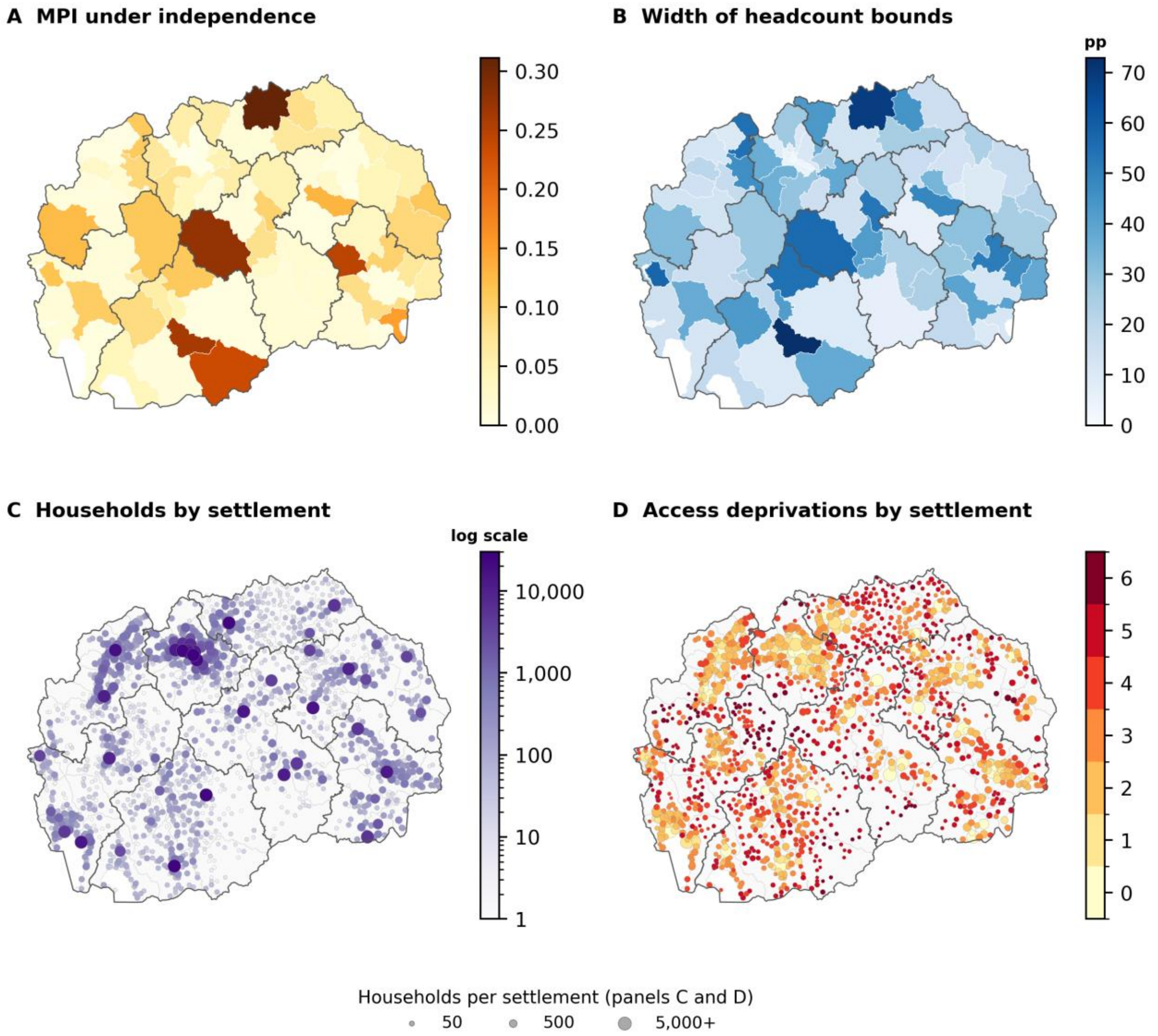

**Figure C1. Poverty estimates, uncertainty, households, and access to services. A Municipal MPI point estimates under independence. B Difference between the upper and lower headcount bounds, in percentage points. C Households per settlement, with color on a logarithmic scale. D Number of access deprivations in each settlement (0–6). Symbol size in C and D indicates the number of households, capped at 5,000. Thick lines mark statistical regions; thin lines mark municipalities. Panels C and D describe observed quantities independently of the assumptions used for the MPI point estimates.**

We assess spatial association across the 80 municipalities using global Moran's I (Moran, 1950). Neighbors share a municipal boundary or corner, and weights are row-standardized so that each municipality's spatial lag is the average value among its neighbors. A one-meter polygon buffer accommodates potential digitizing gaps; for the supplied boundary file, this produces the same neighbor links as exact contiguity. Global inference uses 9,999 random permutations and a two-sided test based on distance from the mean of the permutation distribution.

For the independence-based MPI estimates, Moran's I is 0.0218 (p = 0.6075). The headcount and its lower and upper endpoint maps also have insignificant global results (I = 0.0215, −0.0014, and 0.0785; p = 0.6291, 0.8715, and 0.1905, respectively). The geographic sub-index has I = −0.0090 (p = 0.9526), and the socio-economic sub-index has a weak positive association, I = 0.1097 (p = 0.0858). None is significant at the 5 percent level under these weights and test convention.

We also compute local Moran statistics (LISA; Anselin, 1995), holding the value in each focal municipality fixed and drawing its neighbors' values without replacement from the other 79 municipalities. The 9,999 conditional permutations are compared using the absolute magnitude of the local statistic. High and low refer to values above and below the unweighted municipal mean; local associations describe the focal value and the average of its neighbors. At an unadjusted 5 percent threshold, the MPI flags Rankovtse as High-High (p = 0.0292), Centar as Low-Low (p = 0.0444), and Bitola and Prilep as Low-High (p = 0.0104 and 0.0010). These local findings use unadjusted p-values and should be treated as exploratory. Applying Benjamini–Hochberg adjustment (Benjamini and Hochberg, 1995) within the family of 80 municipal tests leaves no MPI location significant at a 5 percent false discovery rate; the smallest adjusted p-value is 0.08. The reproducibility package reports both the raw p-values and the adjusted values.

These tests describe particular maps. In particular, the MPI point-estimate result is conditional on the independence assumption. Testing the maps formed by all lower headcount bounds and all upper bounds does not test every combination of municipal values allowed by the data, and does not provide bounds on Moran's I. Consequently, the absence of significant global spatial association in the maps examined cannot establish that poverty is spatially unclustered under every admissible dependence structure.